\documentclass[12pt]{article}

\newcommand{\beq}{\begin{equation}}
\newcommand{\eeq}{\end{equation}}
\newcommand{\ket}[1]{\left| {#1} \right>}

\begin{document}

\title{A quantum computer only needs one universe}

\author{A. M. Steane\\
\small Centre for Quantum Computation,\\
\small Department of Atomic and Laser Physics, University of Oxford, \\
\small Clarendon Laboratory, Parks Road, Oxford OX1 3PU, England. }

\date{1 October 2002}

\maketitle

\begin{abstract}
The nature of quantum computation is discussed. It is argued that,
in terms of the amount of information manipulated in a given time,
quantum and classical computation are equally efficient. Quantum
superposition does not permit quantum computers to ``perform many
computations simultaneously'' except in a highly qualified and to
some extent misleading sense. Quantum computation is therefore not
well described by interpretations of quantum mechanics which
invoke the concept of vast numbers of parallel universes. Rather,
entanglement makes available types of computation process which,
while not exponentially larger than classical ones, are
unavailable to classical systems. The essence of quantum
computation is that it uses entanglement to generate and
manipulate a physical representation of the correlations between
logical entities, without the need to completely represent the
logical entities themselves.
\end{abstract}

\maketitle

\vspace{24 pt} The main purpose of this article is to improve our
insight into what is going on in any quantum computation. Although
I have no new quantum algorithms or methods to offer, I hope this
type of contribution may still be of some help towards a better
grasp of quantum computation, and therefore towards future
insights and new methods.

The article has been prompted by the often-quoted, though admittedly vague statement,
``a quantum computer can perform vast numbers of computations simultaneously". I
think this statement is sufficiently misleading that it should have a ``health
warning label'' attached to it, where in this case it is the health of our
insight into quantum computing which we need to guard. The statement is sometimes
used as evidence that quantum physics is best understood in terms of vast
numbers of parallel universes \cite{Bk:DeWitt,Bk:Deutsch}, and
therefore that a laboratory demonstration
of quantum computing is evidence in favour of such interpretations as opposed to
others (though they all make the same predictions). It is not my main purpose
here to bring out the logical difficulties of this, or any other, interpretation
of quantum physics. However, the title ``A quantum computer only needs one
universe'' conveys the conclusion of the present discussion: quantum computers
are not wedded to ``many worlds'' interpretations, not only in terms of the
prediction of the results of experiments, but also in terms of insight into
what is going on within the quantum computational process.

The discussion will consist of a sequence of seven remarks
followed by a proposition. The intention is to point out features
of quantum computing processes which suggest that a better insight
into quantum computing is gained by thinking of it as a small
process which exploits correlations provided by entanglement,
rather than a large process which exploits massive parallelism.
The thesis is that quantum computation offers not a greater amount
of computation in a given system size and time, but rather a more
flexible type of process than is available in classical
computation. The final proposition is a view of what the essence
of this further flexibility is, namely an efficient way to
represent and manipulate correlations.

\section{Seven remarks}

{\em Background to remark 1}. For the purpose of this first
remark, by ``computations'' we mean elementary processing
operations which achieve some given degree of transformation of a
body of information, such as evolving it from one state to an
orthogonal state. It is certainly not self-evident that a quantum
computer does exponentially more computations than a classical
computer of similar size calculating for a similar time, since
there are not exponentially more computational results available.
This follows immediately from Holevo's theorem
\cite{73:HolevoA,Bk:Nielsen} on the capacity of a quantum channel
to transmit classical information. We may deduce that whatever
else may be said about a quantum computer, it does not constitute
many classical information processors. (It is self-evident that it
does constitute one quantum information processor). No one, to my
knowledge, has seriously argued that a quantum computer does
constitute many classical information processors, but informal
statements implying this have been quite common (and I have not
been totally innocent of them).

Furthermore, when a classical
computer simulates the action of a quantum computer, it may need exponentially
more time steps or physical components, but then it also yields exponentially
more information about the final state. Therefore:

\begin{quote}
{\bf Remark 1}. {Quantum computers cannot manipulate classical
information more efficiently than classical ones, and the
total information about the dynamics of a quantum system which can be
obtained by classical computing cannot be obtained more
efficiently by quantum computing.}
\end{quote}

In this sense, the two types of computing are equally efficient.
Nevertheless, the ability of a quantum computer to be focused onto
specific desired results remains highly significant and useful,
for the same reason efficient classical algorithms are significant
compared to inefficient ones.

{\em Background to remark 2}.
Some insight into computational efficiency can be
obtained by examining the difference between an efficient and an inefficient
classical algorithm for the same problem. Take as an example problem that of finding
an item in an ordered list (for example, a name in an alphabetically ordered list).
In order to make the problem capable of being efficient both in space and time,
we assume that elements of the list can be generated by some fast algorithm $f$.
The problem is then equivalent to that of finding the root of a monotonic
function $f(x)$, where $x$ is an integer between zero and $N-1$ (where for
an $n$-bit problem, $N=2^n$).
An efficient algorithm is the binary search (examine $f(N/2)$, and then
according as it is less than or greater than zero, discard the first half
or the second half of the list, and repeat).
An inefficient algorithm is the exhaustive search (examine every element in turn, until
the root is found). If we make a direct comparison between the
binary search and the exhaustive search, forgetting for a moment our
understanding of number theory, then each step of the binary search appears
to accomplish an exponentially large number, of order $N/2$, steps of
exhaustive search. For example if $f(N/2) < 0$ then in one step of the binary
search we have apparently accomplished the $N/2$ `computations'
$f(0) \ne 0,\;f(1) \ne 0;\,f(2) \ne 0,\;\cdots f(N/2) \ne 0$. Actually, of course,
only one of these computations has been carried out: the rest follow by a
process of reasoning, drawing on the definition of number and the statement
that $f(x)$ is monotonic.

\begin{quote}
{\bf Remark 2.}
{The quantity ``amount of computation'' is not
correctly measured by counting the number of steps which would have had to
be accomplished if the computation had been done another way.}
\end{quote}

Therefore, to measure the ``amount of computation" carried out in a
quantum algorithm such as Shor's, it is inappropriate to count the steps
which a classical computer would have needed. A laboratory demonstration
of Shor's algorithm does not constitute proof that huge amounts of
computation have taken place in a small system in a small time---unless
there is a proof to that effect which we have not yet considered.

{\em Background to remark 3.} The ``proof" (or rather, evidence)
usually offered is the presence of processes such as \beq
\sum_{x=0}^{2^n-1} \ket{x} \ket{0} \rightarrow \sum_{x=0}^{2^n-1}
\ket{x} \ket{f(x)},  \label{fx} \eeq in a quantum algorithm such
as Shor's. However, we know that such a process does not
constitute ``evaluation of $N$ values of the function" except in a
highly qualified sense, since upon examining the computer, we will
only be able to learn one value of the function. To that extent
the situation is comparable to the classical binary search, where
when a single function evaluation was carried out, an appearance
of vast numbers of parallel evaluations arose when the algorithm
was looked at from a perspective which lacked insight. Therefore,
it remains open whether the mathematical notation of (\ref{fx}) is
giving a misleading appearance or a good insight into the quantity
of computation.

Let us consider examples of notations which give an impression of
many simultaneous computations, but where we can prove this to be
a false impression. I will propose first an artificial classical
example, and then a more powerful quantum one. Suppose we have a
collection of $n$ compass needles. Each needle can indicate north,
south, east or west, or any other direction. The direction is a
two-component vector which we write with the notation $[\psi]$. We
will use our needles as a simple classical computing device, in
which pointing north represents zero and pointing east represents
one. The vector for north may therefore be conveniently written
$[0]$, and the vector east can be written $[1]$. A state of $n$
needles such as $[0][1][0][1]$ is written $[0101]$. Suppose we
begin with all needles pointing north, and then rotate each needle
to point northeast. This requires $n$ elementary operations, and
performs the process \beq [0] \rightarrow \frac{1}{2^{n/2}}
\sum_{x=0}^{2^n-1} [x].         \label{cc} \eeq Our computer now
``stores all the values of $x$ from $x=0$ to $x=2^n-1$
simultaneously". It has also just ``performed $2^n$ evaluations of
the function $f(x) = x$" in only $n$ steps! Actually, of course,
this computer only ``stores'' all those values in a highly
qualified (and in this case almost useless) sense, and only
``evaluates'' all those function values in a highly qualified
sense, despite the appearance of (\ref{cc}). (More complicated
functions can be evaluated ``in parallel'', by many methods: for
example, by operating a sequence of base-3 logic gates on the
needles, where the three logic states for each gate input and
output are $[0]$, $2^{-1/2}([0]+[1])$ and $[1]$, and then
interpreting the final state of the needles as a superposition of
binary numbers. Of course this is of no practical value, and only
a small family of functions can be treated). This is a good
illustration of the fact that the essential element in quantum (as
contrasted with classical) computing is not superposition but
entanglement. It is the entanglement in the right hand side of
(\ref{fx}) which makes the quantum state computationally useful,
and it is entanglement which is hard to express succinctly in
mathematical notation.

A more powerful example is given by the Gottesman-Knill theorem
\cite{98:GottesmanB,Bk:Nielsen} (I take the statement from
\cite{98:GottesmanB}):
\begin{quote}
{\bf Gottesman-Knill theorem}: {\em Any quantum computer
performing only: a) Clifford group gates, b) measurements of Pauli
group operators, and c) Clifford group operations conditioned on
classical bits, which may be the results of earlier measurements,
can be perfectly simulated in polynomial time on a probabilistic
classical computer.}
\end{quote}
When we recall that the Clifford group contains both Hadamard rotations
and controlled-not gates, we see the strength of this statement. It means
that there exist many quantum algorithms which,
when written down in standard state-vector notation, have an appearance
of multiple parallel computations just as strong as that of (\ref{fx}),
and yet which can be classically simulated efficiently.

\begin{quote}
{\bf Remark 3.} {In view of the fact that it is possible for
mathematical notation to give a false impression of the quantity
of computation represented by a given process, impressions such as the
one contained in (\ref{fx}) do not give a reliable guide to
quantity of computation.}
\end{quote}

Such an impression may merely reflect a
weakness of the mathematical notation, not a profound insight into
what is going on.

To conclude so far,
when a quantum computer is evolved through a process such as (\ref{fx})
it is sometimes stated that the quantum computer `computes' all the function
evaluations $f(x)$. It is then asked, how can this be so, when a classical
computer would need exponentially more time and/or space
to compute all these things?
However, this is a simple case of the same word `compute' being used to mean
two essentially different things, so there is no paradox. The quantum
computer process (\ref{fx}) is being compared to the very
different process
\beq
\ket{0}\ket{1}\ket{2}\cdots\ket{2^n-1}
\rightarrow \ket{f(0)}\ket{f(1)}\ket{f(2)}\cdots\ket{f(2^n-1)}.  \label{pfx}
\eeq
There is no reason why two such thoroughly different processes should require
similar resources.

The argument so far has not produced an indisputable case,
but in view of the remarks made, I would say the burden
of proof lies with those who claim that quantum computation does really
constitute a vast quantity of computation carried out in parallel.
The remaining remarks argue directly against that claim. The aim
of the discussion is not merely to say what quantum computation is
{\em not}, however---I will also argue for an alternative, admittedly incomplete,
view of what it {\em is}.

\begin{quote}
{\bf Remark 4}. {
An $n$-qubit quantum
computer is only sensitive to decoherence to the level $1/\mbox{Poly}(n)$, not
$1/\exp(n)$, in the case that different qubits have independent decoherence.
If the quantum computer were really ``doing $2^n$ computations",
and the result depended on getting a large proportion of them right, then we
would expect it to be sensitive to errors at the level $1/2^n$, which it is
not.}
\end{quote}

I feel this point is so strong that it suffices on its own to rule out the
concept of ``vast parallel computation''.

{\em Background to remark 5}. Quantum computing is now a field
which has reached a modest degree of maturity, but there are still
profound unresolved basic issues, chiefly the nature of
entanglement involving more than two parties, and the general
problem of constructing algorithms which take advantage of quantum
physics. Almost certainly insights into each of these will
contribute to understanding the other. Although most work on
quantum algorithms uses the model of a quantum register with a
network of logic gates, it is well known that other computing
models are possible, for example cellular automata. Most such
models are close cousins of the network model. Recently a new
model was discovered which can be shown to reproduce the results
of the network/register model, but which also can produce
behaviour outside that model. This new model is the `cluster
state' computer, or `one-way computer' discovered by Raussendorf
and Briegel \cite{00:Raussendorf,01:Raussendorf}. The central
elements are the preparation of a special entangled state of many
qubits at the outset of the computation (the cluster state),
followed by appropriately-chosen measurements of single qubits. No
further elements are needed (in particular, no unitary `logic
gates', whether on one or more qubits, are needed, nor are joint
measurements of two or more qubits needed). The choice of
measurements at a given stage depends on the outcome of previous
measurements. It can be shown that this model can be used to
reproduce the action of any quantum network, with similar
resources (qubits and time). However, it can also produce
behaviour which has no natural interpretation in terms of networks
of logic gates. For example, the number of steps (`logical depth')
required to accomplish a desired transformation can be much
smaller (e.g. a constant rather than a logarithm of the input
size), and the temporal ordering of the measurements can be
unrelated to the sequence of gates in a network designed to
accomplish the same algorithm.

The cluster state prepared at the outset is a fixed state which
does not depend on the computation to be performed. The
measurements to be implemented at any time are determined from two
pieces of classical information: a set of angles given by the
algorithm, and the `information flow vector' which is
a classical bit-string of length $2n$ where $n$ is
the size of the input information. This bit-string is updated
depending on the outcomes of the measurements, and half of it gives the
algorithm's output when all measurements are complete.

\begin{quote}
{\bf Remark 5.} The evolution of the cluster-state computer is not
readily or appropriately described as a set of exponentially many
computations going on at once. It is readily described as a
sequence of measurements whose outcomes exhibit correlations
generated by entanglement.
\end{quote}

In order to design an algorithm for this or any other computer, it
is natural to think in terms of classical information in the first
instance, simply because that is the only way we know well. For
example one might start from a network model, analyzed in a
computational basis, and make use of the ``quantum parallelism''
concept of eq. (\ref{fx}). This is certainly one good way to think
about designing algorithms. However, the actual evolution of the
cluster state computer has no ready mapping onto this analysis.
The main features are instead the information flow vector, and the
cluster state whose entanglement slowly disappears as more and
more measurements are made on it. The information being processed
must reside in these two, but the qubits play almost a passive
role, in that they are prepared at the outset in a standard state,
and thereafter simply measured one at a time. Rather than
`performing computations in superposition', the role of the
quantum information is to provide a resource, namely entanglement,
which permits the measurement outcomes to exhibit correlations of
a different nature to those which would be possible with a set of
classical bits.

{\em Background to remark 6.} I have argued that it is not true
that a quantum computer accomplishes a vast number of computations
all at once. A statement which, by contrast, has a clear meaning,
and which I think is more useful, is that a quantum computer can
compute a specific desired result, such as the period of a
function, using much fewer resources than a classical computer
would need. Now, when we examine how it is that some classical
algorithms are more efficient than others, we find (as in the
ordered search example considered above) that the efficient
algorithms do not generate (either temporarily or permanently)
unnecessary subsidiary results. It is natural, therefore, to ask
whether quantum computers out-perform classical ones for the same
reason. In view of the fact that, as I have already argued, the
two types of computer are equally efficient, in terms of quantity
of computations in a given time, this is probably the only
available route for improved efficiency. When we examine an
efficient quantum algorithm such as Shor's, we find that it is
indeed essential to the working of the algorithm that the
evaluations of $f(x)$ in superposition do {\em not} individually
have any subsequent influence on other parts of the universe. If
they did, the resulting entanglement would prevent the algorithm
from working. The algorithm only establishes the correlations,
such as that between $f(x)$ and $f(x + r)$ where $r$ is the
period, not the individual values themselves.

\begin{quote}
{\bf Remark 6.} Whenever one algorithm for a given problem is
substantially more efficient than another, the more efficient
algorithm generates much less extraneous classical information.
\end{quote}

Both memory resources and time must be included when measuring
efficiency. The value of this remark is that it applies uniformly
to classical and to quantum computing, and to their comparison. It
implies that we should understand a gain in computational
efficiency as a given result achieved with less processing, not as
a given result achieved with the same amount of processing but in
parallel.

\begin{quote}
{\bf Remark 7.} {The different ``strands'' or ``paths'' of a
quantum computation, represented by the orthogonal states which at
a given time form, in superposition, the state of the computer
(expressed in some product basis) are not independent, because the
whole evolution must be unitary.}
\end{quote}

This remarks underlines the fact that in a quantum computer a
single process is taking place, not many different ones. One
practical result is that quantum computation cannot give an
efficient algorithm for the unstructured search
problem\cite{97:Bennett}.

\section{Entanglement, superposition and correlations}

It is undisputed that entanglement plays an important role in
quantum computing, though the elucidation of this role is an
ongoing research area. By definition, an entangled state cannot be
written as a product, so if we want to write it down we will have
to write a sum of terms. Owing to the linearity of quantum
mechanics, subsequent unitary operations cause these terms to
evolve independently, and the attraction of the picture of
multiple parallel computations comes from this. However, this
feature is no different from what is observed in the Fourier
analysis of a classical linear electronic circuit. Each Fourier
component of a classical signal will there behave independently of
the others, but it does not give any useful insight to talk of the
different Fourier components as occupying `parallel universes'.

The Fourier example (and others that could be given) emphasizes
that superposition is not in itself the essential ingredient in
quantum computation. Entanglement is, on the other hand, the
essential difference between the states on the right hand side of
equation (\ref{fx}) and of equations (\ref{cc}) and (\ref{pfx}),
and no known efficiency separation between quantum and classical
computation does not involve the exploitation of entanglement for
computational purposes.

I will now put forward an interpretational view of quantum
computing which is in accord with the seven remarks above, and
with what is known about entanglement.

\begin{quote}
{\bf Interpretational view}. A quantum computer can be more
efficient than a classical one at generating some specific
computational results, because quantum entanglement offers a way
to generate and manipulate a physical representation of the
correlations between logical entities, without the need to
completely represent the logical entities themselves.
\end{quote}

The `logical entities' will typically be integers. Thus, for
example, in a set of qubits described by equation (\ref{fx}), the
correlation between $f(x)$ and $x$ is fully represented, but the
values of $f(x)$ are not. For, a measurement of the qubits in the
computational basis will with certainty give a pair of results
such that if one is $x$, the other is $f(x)$, for any $x$ in the
superposition, but it will only with low probability give any
particular $x,\, f(x)$. Furthermore, if the qubits are to be used
in Shor's period-finding algorithm, then the period of the
function which the algorithm extracts is a property of the
correlation between values of $f(x)$, not of any particular value,
and when the algorithm finishes this correlation information is
available, but no physical record remains of any value of $x$
for a given $f(x)$. This is not an insignificant side-effect, because
the absence of a record of any $x$ arises from an
interferometric cancellation which is essential to the success of
the algorithm.

Note also that the interferometric cancellation is only possible
if the terms in the sum are parts of a single entity, i.e. the
single, coherent, state of a system isolated in such a way that it
does not leave `which path' information through entanglement with
other systems. In common with remark 7 above, this emphasizes that
the terms in the superposition do not each have a separate
existence, and therefore should not be described as if they did.

The EPR experiment, in the form as analyzed by Bell, emphasizes
that entanglement leads to a degree of correlation beyond that
which can be explained in terms of local hidden variables. In
order that these correlations are consistent with special
relativity (i.e. that they cannot be used for faster-than-light
signaling) it is necessary that they appear `hidden' in two sets
of measurement results which are random when either set is
examined without the other. This combination of correlation and
randomness is a further example of what I mean by a physical state
which can represent correlation without representing information
about the correlated entities (except in so far as this is
logically necessary to represent information about their
correlation).

To conclude, the basic fact which quantum computers take advantage
of, is that multi-partite entanglement offers a way to produce
some computational results without the need to calculate a lot of
`spectator' results. For example, we can find the period of a
function without calculating all the evaluations of the function;
we can find a specific property of a quantum system (such as an
energy level) without also finding the complete wavefunction; we
can communicate some shared aspect of distributed information
without transmitting as much of the information as we would
otherwise need to.

The impression of vast parallel computation in (\ref{fx}) is a
false impression engendered by an imperfect mathematical notation.
It might be argued that the mathematical notation is the only one
we have, and that it carries a lot of insight into what is going
on in the algorithm. The latter is true, but since we know for a
fact the idea of `vast computation' could only be true in a highly
qualified sense here, and since there is other evidence to suggest
vast computations are not in fact going on, therefore this
impression is merely an artifact of the notation. It is noteworthy
that the very fact that we can write the state using a summation
symbol, rather than writing out all the components laboriously,
indicates that the algorithmic information content of the state is
small.

Entanglement does mean the process is of a subtle type not available to
any classical system. Therefore the computation process, though not
exponentially large, is unavailable to classical computers.

The answer to the question `where does a quantum computer manage
to perform its amazing computations?' is, we conclude, `in the
region of spacetime occupied by the quantum computer'.
Nonetheless, the quantum computer's evolution is a subtle and
powerful process, and one might want to convey this fact by
invoking the image of an `exploration of parallel universes'.
However, since the concept of `parallel universes' implies a
computational power which is not in fact present in quantum
computation, I feel such an image obscures more than it
illuminates.

The right way to describe the efficiency of quantum computation
is, I have argued, that entanglement provides a way to represent
and manipulate correlations directly, rather than indirectly
through a manipulation of the correlated entities.

Finally, if the state vector notation of (\ref{fx}) is imperfect,
then can we think of a notation giving further insight? A more
insightful perspective in many areas of physics is that of
operators rather than states. For example, take the Heisenberg
picture of quantum mechanics, the creation/destruction operator
description of quantum optics, and the stabilizer description of
quantum error correction. We have noted that quantum algorithms
which cannot be efficiently simulated classically exploit
entanglement. A notation which focused on this distinction, i.e.
which treated operations on entanglement measures rather than
state vectors, may give a useful insight.

I acknowledge helpful correspondence with David Deutsch,
Christof Zalka, and Michael P. Frank.
This work was supported by EPSRC and by the Research Training and
Development and Human Potential Programs of the European Union.


\end{document}